\documentstyle[aps,epsf,rotate,subfigure]{revtex}
\font\mb=msbm10
\begin{document}
\draft
\twocolumn[\hsize\textwidth\columnwidth\hsize\csname @twocolumnfalse\endcsname
%
\title{Suppression and Enhancement of Diffusion in Disordered
Dynamical Systems}
\author{R. Klages\cite{em}}
\address{Max Planck Institute for Physics of Complex Systems,
N\"othnitzer Str. 38,  D-01187 Dresden, Germany}
\date{\today}
\maketitle
\begin{abstract}
The impact of quenched disorder on deterministic diffusion in chaotic
dynamical systems is studied. As a simple example, we consider piecewise
linear maps on the line. In computer simulations we find a complicated
scenario of multiple suppression and enhancement of normal diffusion under
variation of the perturbation strength. These results are explained by a
theoretical argument showing that the oscillations emerge as a direct
consequence of the unperturbed diffusion coefficient, which is known to be a
fractal function of a control parameter.
\end{abstract}
]
Recently there has been growing interest in the field of {\em disordered
dynamical systems} thus trying to bring together two at first view very
different directions of research \cite{dds,Rado}: Disordered lattices
exhibiting quenched (static) randomness are considered as a traditional
problem of statistical physics. Hence, in this type of models normal and
anomalous diffusion are studied by probabilistic methods being developed in
the framework of the theory of stochastic processes
\cite{HKW,BoGe90,Kehr98}. On the other hand, diffusion can also be
generated from microscopic chaos in nonlinear equations of motion. This is
well-known as the phenomenon of chaotic diffusion in deterministic dynamical
systems \cite{GNSFK,GF}. Here, methods of dynamical systems theory can be
applied for computing deterministic transport coefficients
\cite{LiLi,GC,Do99}. Disordered dynamical systems represent an interesting
combination of both types of models and provide an ideal opportunity to bring
these two theoretical approaches together. To our knowledge, only very few
cases of such models have been studied so far. Examples include random Lorentz
gases for which Lyapunov exponents have been calculated by means of kinetic
theory and by computer simulations \cite{BDDP}, numerical studies of diffusion
on disordered rough surfaces \cite{PBFH98} and in disordered deterministic
ratchets \cite{PASF00}, as well as numerical and analytical studies of chaotic
maps on the line with quenched disorder \cite{Rado,TsCh}.

In this work we will focus on the most simple case in the latter class of
models, which are piecewise linear maps defined on the unit interval and
acting on the real line by periodic continuation. In case of mixing dynamics,
the unperturbed maps exhibit normal diffusion \cite{GF,ATC,RK,RKdiss,dcrc}. As
first shown in Refs.\ \cite{Rado}, and as further analyzed in Refs.\
\cite{TsCh}, quenched disorder can change the deterministic diffusive dynamics
in these maps profoundly: adding static randomness in form of a local bias
with globally vanishing drift leads to dynamical localization of trajectories
in a complicated potential landscape. On disordered lattices, this effect is
well-known as the Golosov phenomenon \cite{BoGe90,SG} thus reappearing in the
framework of deterministic dynamics. Here we will consider a second
fundamental type of static randomness which is multiplicative and preserves
the local symmetry of the model. Consequently, it does not generate
subdiffusive behavior \cite{supp}. On this basis we study the dependence of
the normal diffusion coefficient on two control parameters, which are the
strength of the unperturbed diffusion coefficient as well as the perturbation
strength \cite{fpm}. Under variation of these two control parameters we find a
complicated scenario for the diffusion coefficient yielding suppression and
enhancement of the strength of diffusion as general features. These findings
will be explained in terms of a simple theoretical approximation, which is
derived from an exact diffusion coefficient formula as obtained within the
theory of stochastic processes.

We first define the unperturbed model by the equation of motion
\begin{equation}
x_{n+1}=M_a(x_n) \quad , \label{eq:eom}
\end{equation}
    
\vspace*{-1.3cm}
\begin{figure}[b]
\epsfxsize=7cm
\centerline{\rotate[r]{\epsfbox{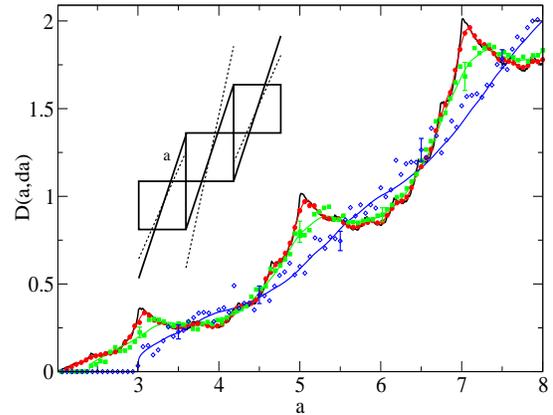}}}
\caption{Diffusion coefficient $D(a,da)$ for the piecewise linear map shown in the
figure. The slope $a$ is perturbed by static disorder of maximum strength $da$
as defined in Eq.\ (\ref{eq:urp}). The bold black line depicts numerically
exact results for the unperturbed diffusion coefficient at $da=0$. Computer
simulation results for $da>0$ are marked with symbols, the corresponding lines
are obtained from the theoretical approximation Eq.\ .(\ref{eq:rka}) The
parameter values are: $da=0.1$ (circles), $da=0.4$ (squares), $da=1.0$
(diamonds).}
\label{fig1}
\end{figure}
where $a\in \hbox{\mb R}$ is a control parameter and $x_n$ is the position of
a point particle at discrete time $n$. $M_a(x)$ is continued periodically
beyond the interval $[-1/2,1/2)$ onto the real line by a lift of degree one,
$M_a(x+1)=M_a(x)+1$. We assume that $M_a(x)$ is anti-symmetric with respect to
$x=0$, $M_a(x)=-M_a(-x)$. The map we study as an example is defined by
$M_a(x)=ax$, where the uniform slope $a$ serves as a control parameter. The
Lyapunov exponent of this map is given by $\lambda(a)=\ln a$ implying that for
$a>1$ the dynamics is chaotic. We now modify this system by adding a random
variable $\Delta a(i)$, $i\in \hbox{\mb Z}$, to the slope on each interval
$[i-1/2,i+1/2)$ yielding $M_a^{(i)}(x)=(a+\Delta a(i))x$. We assume that the
random variables $\Delta a(i)$ are independent and identically distributed
according to a distribution $\chi_{da}(\Delta a)$, where $da$ is again a
control parameter. In the following we will consider two different types of
such distributions, namely random variables distributed uniformly over an
interval of size $[-da,da]$,
\begin{equation}
\chi_{da}(\Delta a)=(\Theta(da+\Delta a)\Theta(da-\Delta a))/(2da) \quad , \label{eq:urp}
\end{equation}
and dichotomous or $\delta$-distributed random variables,
\begin{equation}
\chi_{da}(\Delta a)=(\delta(da+\Delta a)+\delta(da-\Delta a))/2 \quad . \label{eq:drp}
\end{equation}
Since $|\Delta a|\le da$, we denote $da$ as the perturbation strength. As an
example, we sketch in Fig.\ \ref{fig1} the map resulting from the disorder of
Eq.\ (\ref{eq:urp}) as applied to the slope $a=3$. In the absence of any bias,
the diffusion coefficient is defined as
$D(a,da)=\lim_{n\to\infty}<x_n^2>/(2n)$, where the brackets denote an ensemble
average over moving particles for a given configuration of disorder. An
additional disorder average is not necessary because of self-averaging. Note
that for locally symmetric quenched disorder and $(a-da)>2$ there is no
physical mechanism leading to infinitely high reflecting barriers as they are
responsible for Golosov localization \cite{Rado}. Thus diffusion must be
normal, as is confirmed by computer simulations. Hence, the central question
is what happens to the parameter-dependent diffusion coefficient $D(a,da)$
under variation of the two control parameters $a$ and $da$.

For the unperturbed case $da=0$ the diffusion coefficient has been computed
numerically exactly in Refs.\ \cite{RK,RKdiss}. There it has been shown that
$D(a,0)$ is a fractal funtion of the slope $a$ as a control parameter. This
function is depicted in Fig.\ \ref{fig1}, as well as results from computer
simulations for different values of the perturbation strength $da$
\cite{cs}. As expected, the fractal structure smoothes out by
increasing $da$. However, it is remarkable that even for large perturbation
strength $da$ oscillations are still visible as a function of $a$ indicating
that the original irregularities are very robust against uniform quenched
disorder.

Before we proceed to more detailed simulation results we briefly repeat what
is known for diffusion in lattice models with random barriers
\cite{HKW,BoGe90,Kehr98,ZDL}. In the most simple version, the quenched
disorder is defined on a one-dimensional periodic lattice with transition
rates between neighboring sites $i$ and $i+1$ having the symmetry
$\Gamma_{i,i+1}=\Gamma_{i+1,i}\equiv\Gamma_k$ for a given random distribution
of $\Gamma_k$. In this situation an exact expression for the diffusion
coefficient has been derived reading \cite{Kehr98,ZDL}
\begin{equation}
D=\{1/\Gamma\}^{-1} l^2 \quad , \label{eq:sdc}
\end{equation}
with the brackets defining the disorder average
$\{1/\Gamma\}=1/N\sum_{k=0}^N1/\Gamma_k$ at chain length $N$, and for a
distance $l$ between sites. The double-inverse accounts for the physical
significance of the fact that the highest barriers dominate diffusion in
$d=1$. In other words, the existence of rates with $\Gamma_k\to0$ naturally
leads to a vanishing diffusion coefficient $D\to0$. This scenario is
translated to the map under consideration as follows: Eq.\ (\ref{eq:eom}) can
be understood as a time-discrete Langevin equation, $x_{n+1}=x_n-\partial
V/\partial x(x_n)$
\cite{Rado,GF}.  In case of quenched disorder, and under proper integration of
$M_a^{(i)}(x)$, the resulting potential $V(x)$ corresponds to the one of a
random barrier model in which the perturbation strength $da$ determines the
highest barriers. For the disordered map one may thus naively expect
suppression of diffusion reflected in $D(a,da)$ being a monotoneously
decreasing function of $da$.\\[-10ex]
\begin{figure}[b]
\begin{center}
\epsfxsize=6cm
\subfigure{\rotate[r]{\epsfbox{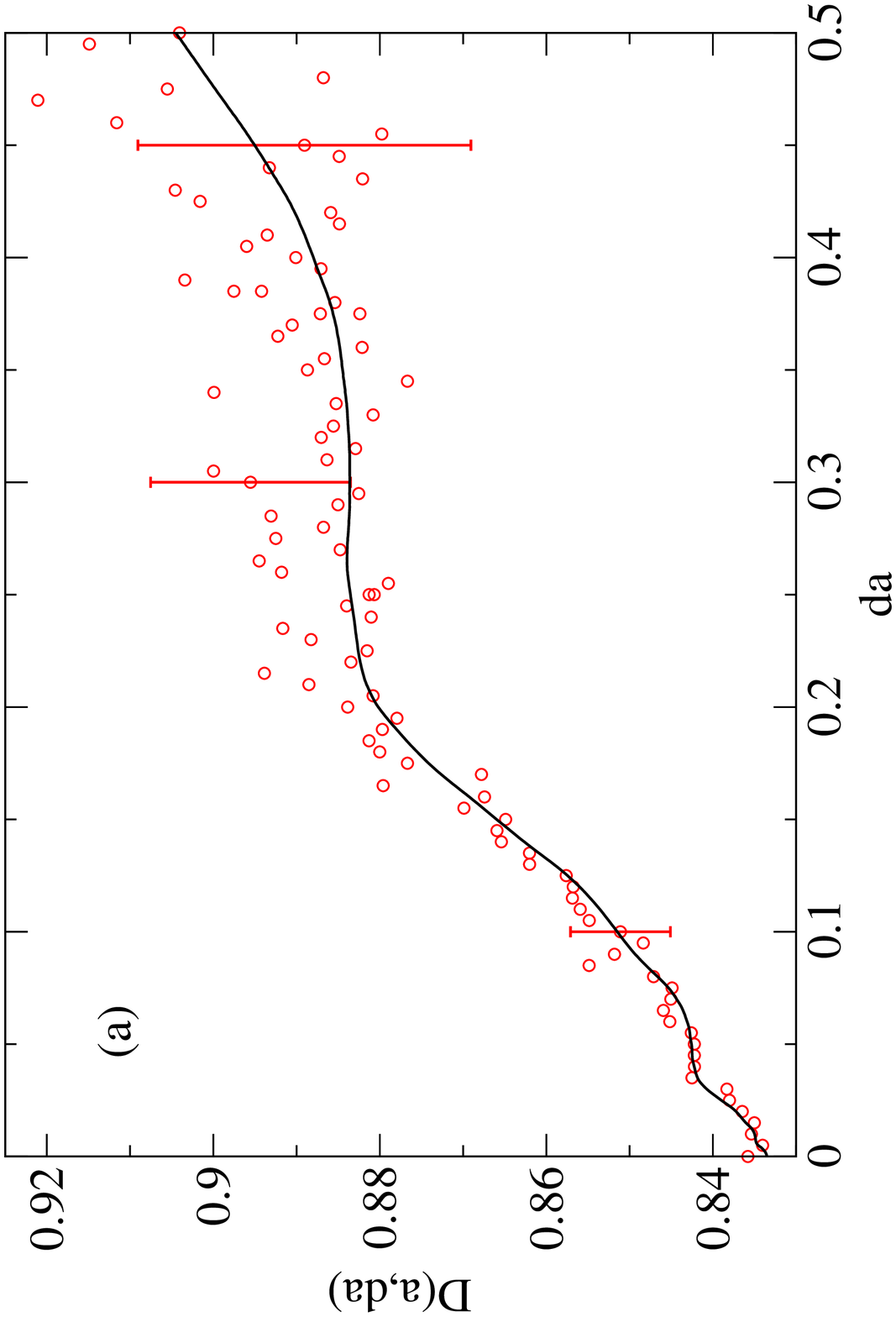}}}\\[-10ex]
\epsfxsize=6cm
\subfigure{\rotate[r]{\epsfbox{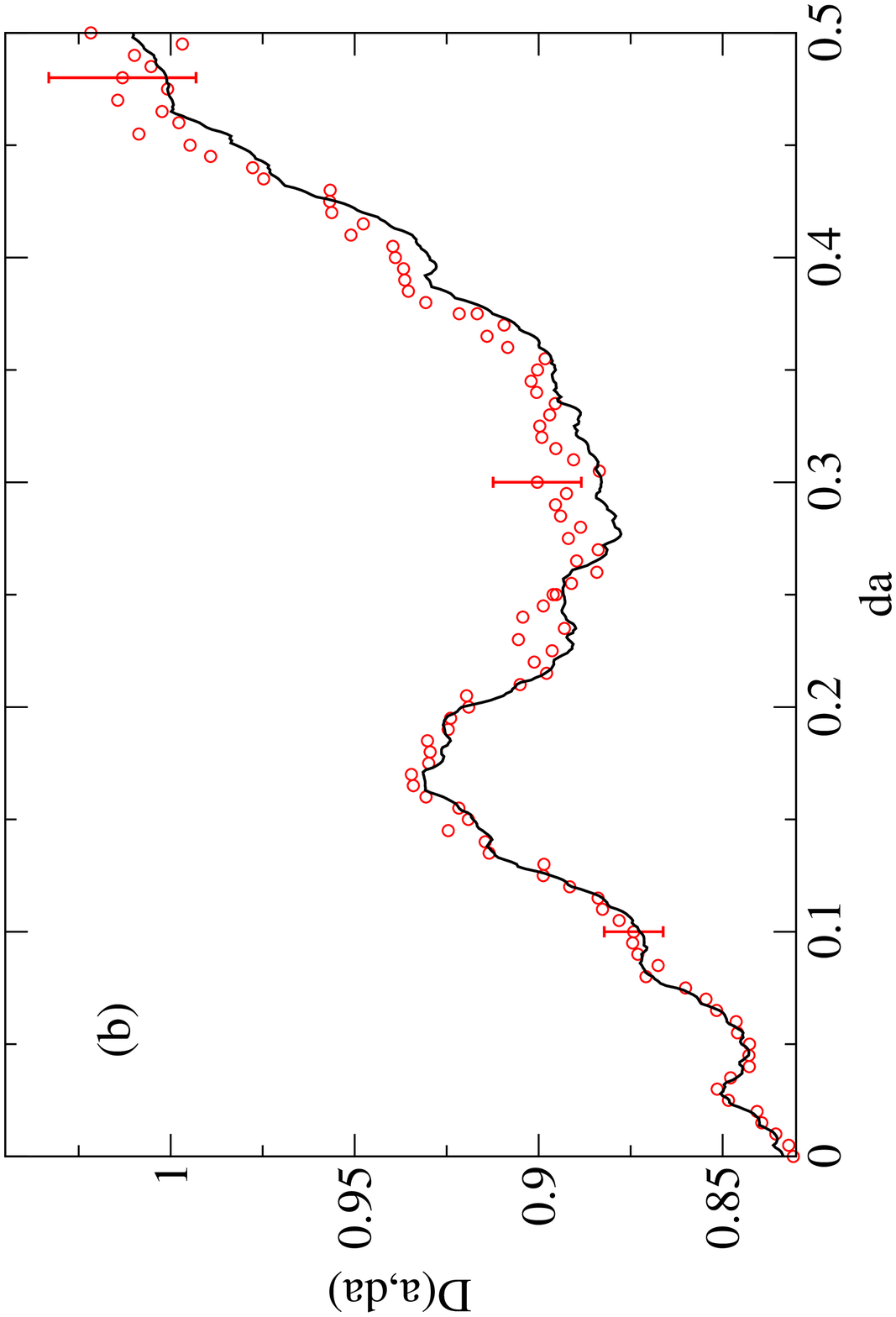}}}
\end{center}
\vspace*{-0.6cm}
\caption{Diffusion coefficient $D(a,da)$ as a function of the perturbation strength $da$
at slope $a=6$: (a) disorder distribution Eq.\ (\ref{eq:urp}), (b) disorder
distribution Eq.\ (\ref{eq:drp}). The circles represent results from computer
simulations, the lines are obtained from the theoretical approximation Eq.\
(\ref{eq:rka}).}
\label{fig2}
\end{figure}
To check this hypothesis, we choose fixed values of $a$ corresponding to the
two extreme situations of sitting at a local maximum or minimum, respectively,
of the unperturbed $D(a,0)$ in Fig.\ \ref{fig1}. We first focus on the local
minimum at $a=6$ for $da\le0.5$ with uniform (Fig.\ \ref{fig2} (a)) as well as
dichotomous (Fig.\ \ref{fig2} (b)) disorder. In sharp contrast to the
stochastic argument outlined above, in both cases we observe enhancement of
diffusion as a function of $da$.  Moreover, this enhancement does not appear
in form of a simple functional dependence on $da$: In (a), smoothed-out
oscillations are visible on smaller scales, whereas in (b) the resulting
function is clearly non-monotoneous and wildly fluctuating exhibiting multiple
suppression and enhancement in different parameter regions. What happens on
larger scales of $da$ is depicted in Fig.\ \ref{fig3}, where we only show
results for dichotomous disorder. As can be seen in Fig.\ \ref{fig3} (b),
choosing $a$ at a local maximum of $D(a,0)$ leads to suppression of diffusion
for $da<1.0$, whereas a local minimum generates enhancement in the same
parameter region of $da$. However, in both cases the diffusion coefficient
decreases on a larger scale thus recovering qualitative agreement with the
expectation from stochastic theory. Indeed, for $(a-da)\to2$ barriers are
formed which a particle cannot cross anymore implying the existence of
localization \cite{undis}.\\[-10ex]
\begin{figure}
\begin{center}
\epsfxsize=6cm
\subfigure{\rotate[r]{\epsfbox{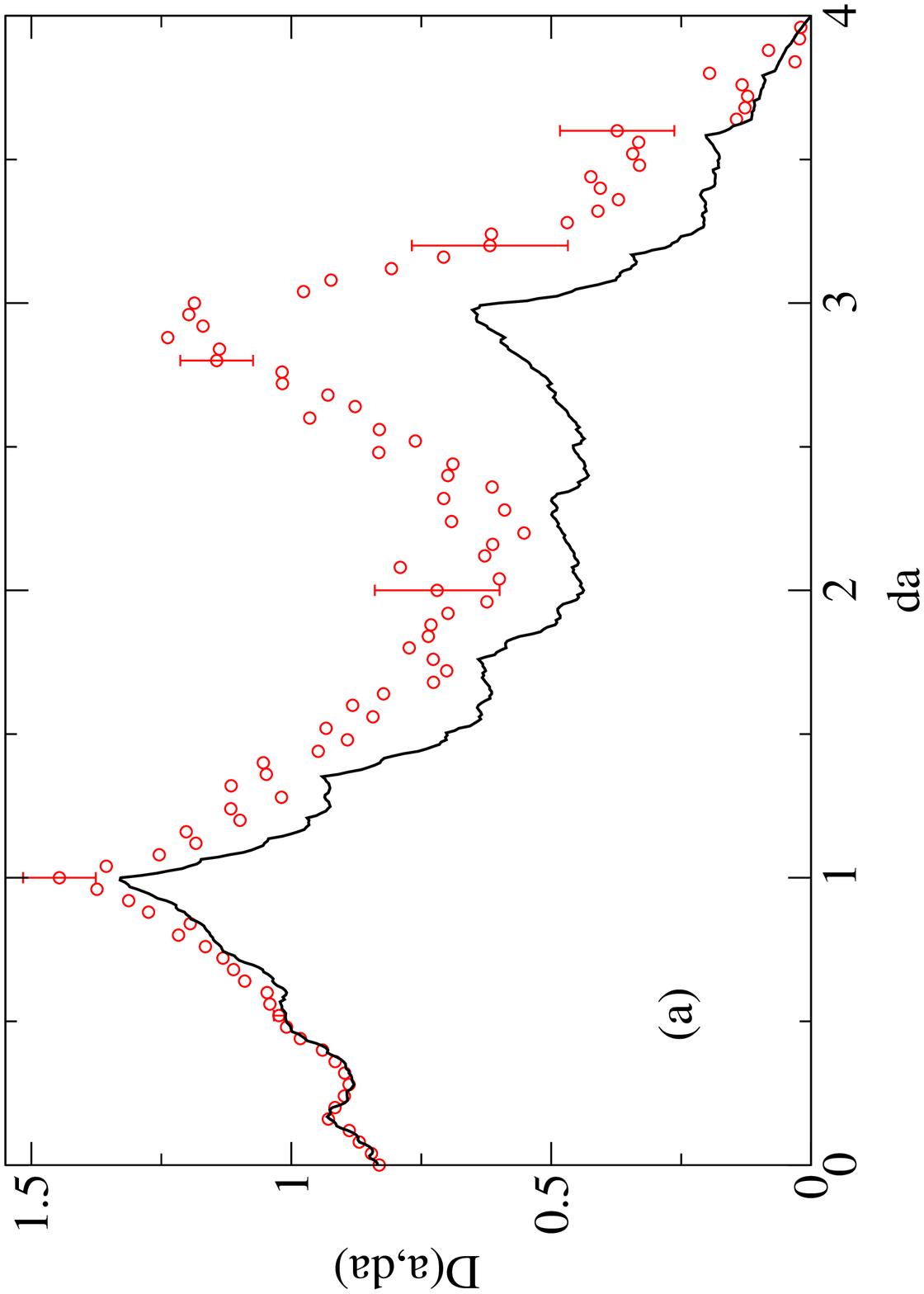}}}\\[-9ex]
\epsfxsize=6cm
\subfigure{\rotate[r]{\epsfbox{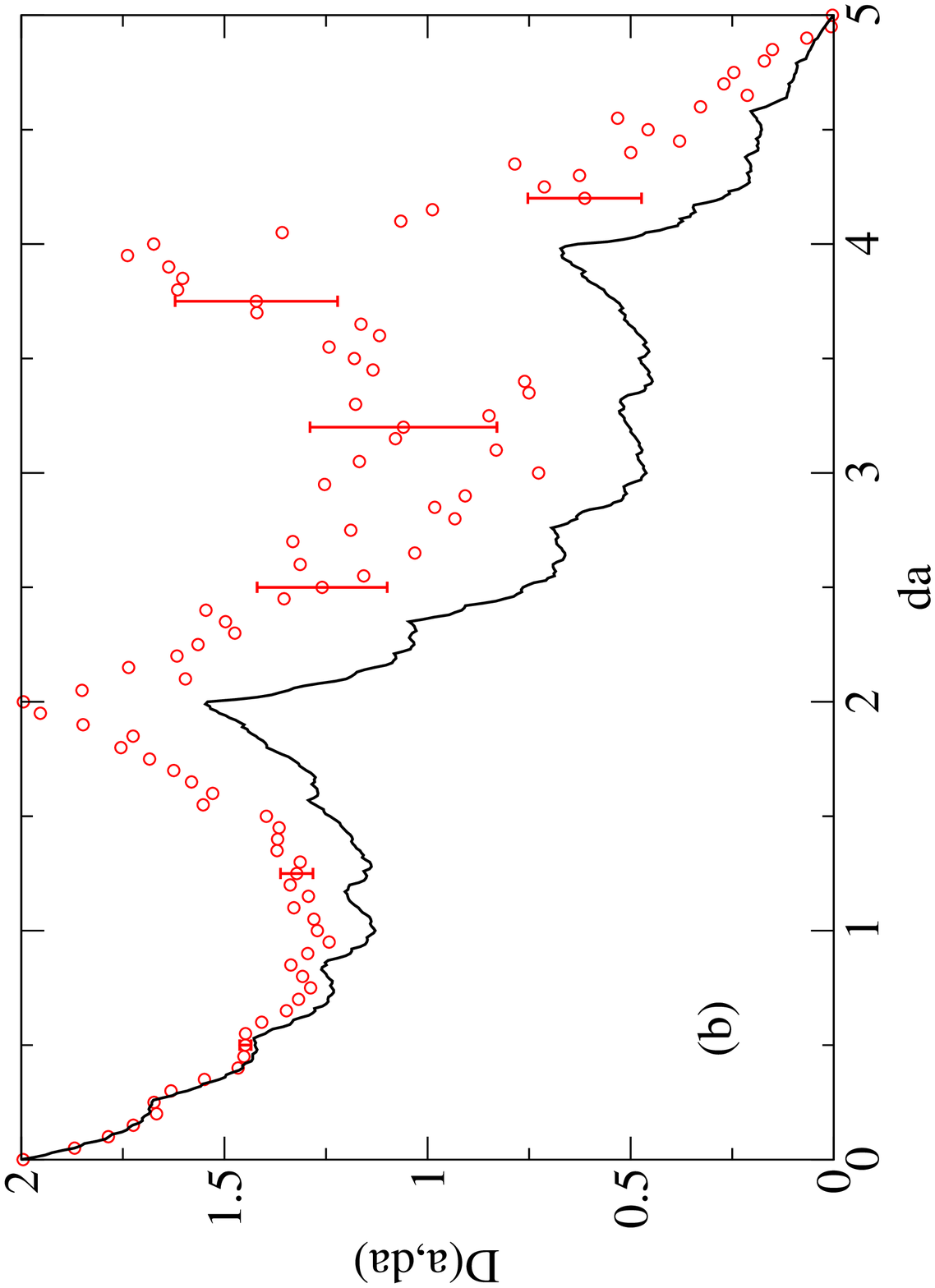}}}
\end{center}
\vspace*{-0.6cm}
\caption{Diffusion coefficient $D(a,da)$ as a function of the perturbation strength $da$
for dichotomous disorder Eq.\ (\ref{eq:drp}) at two different slopes $a$: (a)
$a=6$, (b) $a=7$. The circles represent results from computer simulations, the
lines are obtained from the theoretical approximation Eq.\ (\ref{eq:rka}).}
\label{fig3}
\end{figure}
To theoretically explain these simulation results, we modify Eq.\
(\ref{eq:sdc}) in a straightforward way such that it can be applied to our
disordered deterministic map under consideration. We first note that for
uniform transition rates $\Gamma_k=const.$ it is $D(\Gamma_k,l)=\Gamma_k
l^2$. Using this familiar expression for the random walk diffusion coefficient
on the unperturbed lattice we rewrite Eq.\ (\ref{eq:sdc}) as
$D=\{1/D(\Gamma_k,l)\}^{-1}$. In case of our map, the transition rates and the
distance between sites are both somewhat combined in the action of the slope
$a$ as a control parameter. Therefore, the unperturbed diffusion coefficient
is correctly rewritten by replacing $D(\Gamma_k,l)\equiv D(a)$. Up to this
point we performed purely formal manipulations. The key question is now what
function shall be used for the parameter-dependent diffusion coefficient
$D(a)$ in case of deterministic dynamics. Here we propose {\em to identify the
function $D(a)$ with the exact, unperturbed deterministic diffusion
coefficient} previously defined as $D(a,0)$. Providing this information, the
exact formula Eq.\ (\ref{eq:sdc}) for stochastic dynamics becomes an {\em
approximation} which can straightforwardly be applied to deterministic
dynamics in disordered systems. If the disorder distributions
$\chi_{da}(\Delta a)$ are bounded by the perturbation strength $da$, and if we
go into the continuum limit for the random variable, our final result reads
\begin{equation}
D_{app}(a,da)=\left[\int_{-da}^{da} d(\Delta a)\: \frac{\chi_{da}(\Delta
a)}{D(a+\Delta a)}\right]^{-1} \quad . \label{eq:rka}
\end{equation}
This expression represents the central formula of our letter. The results
obtained from it are depicted in Figs.\ \ref{fig1} to \ref{fig3} in form of
lines. For small enough $da$ the agreement between theory and simulations is
excellent, both for dichotomous as well as for uniform disorder. Surprisingly,
for dichotomous disorder the theory even correctly predicts the oscillations
for larger $da$, although there are clear quantitative deviations.

We now show that this formula provides a simple physical explanation for the
complicated dependence of the diffusion coefficient on the perturbation
strength. In case of $da\to0$, Taylor expansion leads to
\begin{equation}
D_{app}(a,da)=\int_{-da}^{da} d(\Delta a)\: \chi_{da}(\Delta a)D(a+\Delta a)
\quad . \label{eq:dapp}
\end{equation}
We remark that, alternatively, Eq.\ (\ref{eq:dapp}) can be proven for small
enough $da$ starting from the precise definition of the diffusion coefficient,
and without employing Eq.\ (\ref{eq:rka}) \cite{tbp}. In this limit the
perturbed diffusion coefficient reduces to an average of the exact diffusion
coefficient over the neighborhood in the parameter interval $[a-da,a+da]$
weighted by the respective disorder distribution $\chi_{da}(\Delta
a)$. Consequently, if $a$ is chosen at a local minimum the result must be
enhancement of diffusion by increasing the perturbation strength, and
suppression at a local maximum, respectively \cite{rkconj}. On these grounds
it is also clear that the fractal parameter dependence of $D(a,0)$ must
reappear in the perturbed diffusion coefficient thus leading to multiple
suppression and enhancement on fine scales. In case of dichotomous noise, and
if Eq.\ (\ref{eq:dapp}) holds, $D(a,da)$ must be a superposition of two
fractals resulting in a new fractal. In case of uniform noise the fractality
smoothes out, but is still responsible for the oscillations on a fine scale.

We conclude with a few remarks: (a) It would be important to have a proof of
Eq.\ (\ref{eq:rka}) for dynamical systems, as well as to obtain higher order
corrections for explaining the deviations between simulation and theory as
visible in Fig.\ \ref{fig3}.  (b) Our results might be important to understand
diffusion on a stepped surface with a disordered arrangement of
Ehrlich-Schwoebel barriers: As shown in Ref.\ \cite{MWK98}, this problem can
be modeled by a potential relief consisting of a combination of random traps
and random barriers. Our map provides a deterministic generalization of such a
model and particularly enables to study the impact of long-range memory
effects on surface diffusion.  (c) One could think of applying our approach to
systems such as the ones studied in Refs.\ \cite{PBFH98,PASF00}, or to the
periodic Lorentz gas \cite{LiLi,GC,Do99}, which is a model being close to
experiments on antidot lattices \cite{Weis91}. Knowing the density-dependent
diffusion coefficient in the unperturbed case \cite{KlDe00} leads us to
predicting local and global suppression of the diffusion coefficient in this
model in case of static density fluctuations.

We wish to thank Prof.\ K.W.\ Kehr for pointing out to us the existence of
Eq.\ (\ref{eq:rka}), unfortunately, he did not live to see of how much use his
hint was for the present work. The author is also grateful to G. Radons, H.\
van Beijeren, J.R.\ Dorfman, and T.\ Tel for helpful discussions.\\[-4ex]


\begin{thebibliography}{10}
\bibitem{supp}
Note that what is called suppression of diffusion in Ref. \cite{Rado} is a
mechanism leading to subdiffusion. Here we denote with suppression and
enhancement of diffusion the variation of the strength of the normal diffusion
coefficient.
\bibitem{fpm}
We remark that in previous work on disordered maps only fixed specific
parameter values have been chosen where the reduced map corresponds to a
Bernoulli shift \cite{Rado,TsCh}.
\bibitem{cs}
For each point, a maximum of 1,000,000 particles has been iterated up to
50,000 time steps each. Some representative error bars are included in all
figures.
\bibitem{undis}
Simulations have also been carried out for uniform disorder, see Eq.\
(\ref{eq:urp}). In analogy to Figs.\ \ref{fig2} and \ref{fig3}, suppression
for $a=7$ up to $da<1$ and global suppression for $a=6$ and $a=7$ over the
full range of $da$ have been obtained as well, however, with much larger error
bars; these results will be presented elsewhere \cite{tbp}.
\bibitem{rkconj}
Based on this heuristical argument, these phenomena have already been
conjectured in Ref.\ \cite{RKdiss}, see p.\ 65.
\end{thebibliography}

\end{document}